\documentclass[aps,prd,showpacs,
amsmath,amssymb,floatfix,nofootinbib,english]
{revtex4-1} 
\usepackage[T1]{fontenc}
\usepackage[latin9]{inputenc}
\usepackage{textcomp}
\usepackage{amssymb}
\usepackage{epsfig}
\usepackage{graphicx}
\usepackage{esint}
\usepackage{comment}
\usepackage{color}
\PassOptionsToPackage{normalem}{ulem}
\usepackage{ulem}
\usepackage{babel}
\usepackage{slashed}
\newcommand{\be}{\begin{equation}}
\newcommand{\ee}{\end{equation}}
\newcommand{\bea}{\begin{eqnarray}}
\newcommand{\eea}{\end{eqnarray}}

\newcommand{\ba}{\begin{array}}
\newcommand{\ea}{\end{array}}
\newcommand{\bd}{\begin{displaymath}}
\newcommand{\ed}{\end{displaymath}}

\def\gev{{\rm \,Ge\kern-0.125em V}}
\def\tev{{\rm \,Te\kern-0.125em V}}

{
{

\def\th13 {\theta_{13}}

\begin{document}

\title{Inflation, reheating, leptogenesis and bounds on soft supersymmetry breaking parameters}

\author{Imtiyaz Ahmad Bhat}
\email{imtiyaz@ctp-jamia.res.in}

\author{Girish Kumar Chakravarty}
\email{girish@ctp-jamia.res.in}

\author{Rathin Adhikari}
\email{rathin@ctp-jamia.res.in}

\affiliation{Centre for Theoretical Physics, Jamia Millia Islamia, New Delhi, India}

\begin{abstract}
In the no-scale supergravity with Type-I Seesaw model of Non-minimal supersymmetric standard model (NMSSM), we have analysed inflation, reheating and leptogenesis. A no-scale supergravity realization of Starobinsky model of inflation in simple Wess-Zumino model have been shown earlier by Ellis et al. Here we show a no-scale supergravity realization of Starobinsky model of inflation in Type-I Seesaw framework of NMSSM. In this framework an appropriate choice of no-scale K\"ahler potential results in Starobinsky like plateau inflation along a Higgs-sneutrino $D$-flat direction consistent with the CMB observations. In leptogenesis, the soft-breaking trilinear and bilinear terms play important role. Using conditions for non-thermal contribution to $CP$ asymmetry and successful leptogenesis together with the appropriate reheating at the end of inflation, we have obtained important constraints on the soft supersymmetry breaking parameters.
\end{abstract} 


\maketitle 

\section{Introduction}
\label{sec:dm}

 An epoch of rapid exponential expansion is needed to solve the horizon and flatness problem and this is dubbed as the inflation  \cite{Guth,Linde,Guth1,Linde1,Steinhardt} (For non-standard inflationary scenario like brane world inflation readers are suggested to go through {\cite{Randall:1999ee,Randall:1999vf,Maartens:2010ar,Langlois:2002bb,Adhikari:2020xcg}). For short review of Inflation and cosmological perturbation theory and supergravity inflation model building, we refer the reader to articles \cite{Chakravarty:2016set,Cerdeno:1998hs}. The scalar field which could drive inflation is not usually connected to particle physics. This is because the inflation is required to satisfy the slow-roll condition and then the couplings associated with the scalar field does not in general match with the couplings of scalar fields in particle physics. The low upper bound on the tensor-to-scalar ratio by Planck-2018~\cite{Akrami:2018odb} and BICEP2+Keck Array~\cite{Ade:2018gkx} rules out the standard particle physics models with quartic and quadratic potentials. Still in the particle physics Standanrd model context, there exists many appealing models, such as, a successful Higgs inflation scenario has been discussed in \cite{Bezrukov:2007ep} but it suffers from unitarity violation problem. For other Higgs as well as sneutrino-Higgs inflationary scenarios in supergravity embedding of MSSM and NMSSM, we refer the reader to articles \cite{Einhorn:2009bh,Ferrara:2010yw,Lee:2010hj,Chakravarty:2016avd,Dubinin:2017irg}. Another important surviving model is the Starobinsky $R + R^2$ model~\cite{Starobinsky} which predicts a very low tensor-to-scalar ratio of order $10^{-3}$. In the no-scale supergravity framework it was shown by Ellis et al~\cite{Ellis:2013xoa} that a no-scale K\"ahler potential along with a simple Wess-Zumino model superpotential results in a Starobinsky type plateau inflation. A power-law models of Starobinsky inflation $R+R^{\beta}$ in no-scale SUGRA, which can accommodate larger values of tensor-to-scalar ratio than $r\sim10^{-3}$, has been derived and discussed in \cite{Chakravarty:2014yda}. There has been a lot of work done in no-scale supergravity inflation from $F$-term \cite{AlvarezGaume:2010rt,Ellis:2014dxa,Chakravarty:2016fin} and $D$-term scalar potentials \cite{Ferrara:2013rsa,Farakos:2013cqa,Ferrara:2014rya,Nakayama:2016eqv,Chakravarty:2017hcy}.

 In this article we study the Starobinsky-like scenario in the Type-I Seesaw model. The earlier work on SUGRA inflation in Type-I Seesaw has been studied in \cite{Garg:2017tds}.
To get small active neutrino masses around eV scale, if one try to avoid too small Yukawa couplings of neutrinos, the seesaw mechanism with heavy right handed neutrinos is preferred. In the Non-minimal extension of Minimal Supersymmetric Standard Model (NMSSM), there are singlet heavy right handed neutrinos which could explain the light neutrino mass with not too small Yukawa couplings of neutrinos with scalars. The sneutrinos which are scalars and superpartner  of these heavy right handed neutrinos, could play the role of inflation and then the slow roll condition of inflation potential could be addressed. There is another cosmological problem: the observed baryonic asymmetry of the universe over antibaryons. One suitable mechanism which could address this problem is baryogenesis via leptogenesis through sphaleron transition. In leptogenesis, the leptonic asymmetry could be created from the out of equilibrium decays of heavy right-handed neutrinos and sneutrinos. In general, for successful leptogenesis with heavy right handed neutrinos requires the mass of such heavy neutrinos to be above $10^9$ GeV. However, after inflation, large number of massive gravitinos may be produced and their decays could modify the abundance of light elements as observed. To avoid this problem, the reheating temperature is required to be within $10^6$ to $10^9$ GeV. This gravitino problem could be circumvented if one considers soft-leptogenesis in which soft supersymmerty breaking terms in the Lagrangian are considered. In presence of these terms, heavy sneutrino and anti-sneutrino in the same supermultiplet will mix. Leptogenesis could occur below $10^9$ GeV in which the asymmetry is generated from the decay of these mixed eigenstates and the gravitino problem could be avoided. Thus heavy right handed neutrinos and sneutrinos could play role in successful inflation, reheating, leptogenesis as well as in the formation of light neutrino mass. In this work, we try to correlate all these issues and  find out what kind of conditions on various parameters including Yukawa couplings of heavy right handed neutrinos and soft supersymmetry breaking parameters, are required to be satisfied.

In leptogenesis the sneutrino and antisneutrino mixing was initially discussed in \cite{Hirsch:1997vz,Grossman:1997is,Chun:2001mm}. However, to get the required leptonic asymmetry from such mixed states, it was shown later \cite{D'Ambrosio:2003wy,Grossman:2003jv}   that the finite temperature effect associated with phase space and the statistical factors associated with fermion and boson final states, are to be taken into account. 
Later on, it was shown in \cite{Adhikari:2015ysa} that for generic trilinear soft supersymmetry breaking parameter $A$, the sufficient leptonic asymmetry could be produced non-thermally at $T=0$. $CP$ asymmetry could be possible near resonance for which the other soft breaking parameter $B$ is below supersymmetry breaking scale $m_{susy}$  or at away from resonance for which even $B \sim m_{susy} $ could be possible. The higher order correction  for the decays of mixed states 
may come from vertex corrections and self-energy corrections (mixing). However, with very small mass splitting of mixed eigenstates, the $CP$ asymmetry corresponding to higher order diagrams with self energy corrections dominates over vertex corrections. However, one important criteria of the higher order diagrams for the lepton number violating decays to  get $CP$ asymmetry non-thermally is that there should be non-zero lepton number violation on the right of the cuts \cite{raghu}. Some self energy diagrams satisfy this criteria and will contribute significantly to non-zero $CP$ asymmetry at $T=0$ for non-thermal leptogenesis. 

In our work, we have considered this scenario of leptogenesis with generic $A$ parameters in further details by taking into  account various scattering processes which could washout the generated leptonic asymmetry from the decays. Using Boltzman equations we have obtained the leptonic asymmetry and the freeze-out temperature depending on the choices of various soft susy-breaking parameters.

The paper is organized as follows. In Section~\ref{sec:model}, we have derived the model in sugra framework and discussed in detail the $F$-term SUGRA inflation model in NMSSM fields where heavy right handed neutrino fields are present. In Section~\ref{sec:NA}, for Starobinsky like plateau potentials in Type-I Seesaw setup, various conditions on different parameters have been discussed and numerical analysis for the CMB observables have been done. In section \ref{sec:lepto}, there is discussion on leptogenesis from the mixed eigenstates of sneutrino antisneutrino mixing. From the decay of such states leptonic asymmetry is shown to be generated and constraints on $A$ and $B$ parameters is obtained for non-thermal leptogenesis with generic $A$ coupling. Particularly, in section IIIA, $CP$ asymmetry, in section IIIB, details of Boltzman equations and in section \ref{sec:results}, there is short discussion on constraints of soft susy breaking parameters from neutrino mass related with successful inflation, requirement of reheating, non-thermal condition in leptogenesis, out of equilibrium conditions
of the decays of sneutrinos.
 In section \ref{sec:con}, some concluding remarks have been given. 

\section{Supergravity Inflation in Type-I Seesaw}
\label{sec:model}
\subsection{The model}
 In this model we study inflation and
leptogenesis with two Higgs doublets and  left-handed lepton doublet $\ell_{\alpha}$. We have assumed that only two heavier singlet right-handed Majorana neutrinos 
$N_{i=2,3}$ plays the role in inflation. Inflation is driven by inflaton field which is a linear combination of Higgs and sneutrinos.
Sneutrinos are scalar supersymmetric partner of Dirac and Majorana neutrinos. We study a class of model where
inflaton superfield and a string theory moduli field $T$ , which stabilizes the cosmological vacuum, are embedded in no-scale supergravity sector defined by $SU(2,1)/SU(2) \times U(1)$ symmetry. 
We consider  Ka\"hler potential corresponding to no-scale supergravity and the superpotential in NMSSM superfields which has right handed neutrino fields as required for Type-I seesaw-model for light neutrino masses and write those as: 

\begin{equation}
K = -3 \log \left[T +T^{*} -\frac{1}{3}\left(H_{u}^{\dagger} H_{u}+H_{d}^{\dagger} H_{d} + \hat\ell_{\alpha}^{\dagger} \hat \ell_{\alpha} + \hat N_{i}^{*} \hat N_{i}\right)\right]
\end{equation}
\begin{equation}
W = \mu H_{u}.H_{d} + \frac{1}{2}M_i\hat{N_i^{c}}\hat{N_i^{c}}
+Y_{i\alpha}\hat{N_i^{c}}\hat{\ell}_{\alpha}\hat{H}_{u},
\label{eq:Spot}
\end{equation}
where,
\begin{equation}
 H_{u} =\begin{pmatrix}
  \phi^{+}_{u}  \\
  \phi^{0}_{u}
 \end{pmatrix}\,,~~~
 H_{d} =\begin{pmatrix}
  \phi^{0}_{d}  \\
  \phi^{-}_{d}
 \end{pmatrix}\,,~~~
\ell =\begin{pmatrix}
  \phi_{\nu}\\
  \phi_{e}
 \end{pmatrix}\,,~~~
\end{equation} 
Here, $N_i$, $\ell_\alpha$ and $H_{u,d}$ are the chiral superfields for the
right-handed neutrinos, the left-handed lepton doublets and the up and down type Higgs respectively. In the superpotential, the $SU(2)_{L}$ contraction between $\hat{\ell}_{\alpha}$ 
and $\hat{H}_{u}$ is left implicit.

$M_i$ are the different diagonal mass matrix elements, in which $i = 1, 2, 3$ are the diagonal elements $(ii)$  in the heavy right handed neutrino block of the seesaw mass neutrino mass matrix and $M_2$ and $M_3$ have been assumed to be heavier than $M_1$.  In $\ell_\alpha$,  $\alpha=1,2,3$ are the flavor indices. For simplicity, we are assuming all non-diagonal elements of $Y_{i\alpha}$ are very small in comparison to diagonal elements. All neutral lepton fields are equivalent with their zero mass and all neutral slepton fields are equivalent with their degenerate mass. 

In supergravity, the scalar potential depends upon the K\"ahler function $G(\phi_{i},\phi^{*}_{i})$ given in terms of  K\"ahler potential $K(\phi_{i},\phi^{*}_{i})$ and a holomorphic superpotential $W(\phi_{i})$ as $ G(\phi_{i},\phi^{*}_{i}) \equiv K(\phi_{i},\phi^{*}_{i}) + \ln W(\phi_{i}) +\ln W^{\ast}(\phi^{*}_{i})$, where $\phi_{i}$ are the chiral scalar superfields. 
In $\mathcal{D}=4$, $\mathcal{N}=1$ supergravity, the total tree-level supergravity scalar potential is given as the sum of $F$-term and $D$-term potentials given by
\begin{equation}
V_{F}=e^{G}\left[\frac{\partial G}{\partial \phi^{i}} K^{i}_{j*} \frac{\partial G}{\partial \phi^{*}_{j}} - 3 \right] \label{LV}
\ee
and
\begin{equation}
V_D= \frac{1}{2}\left[\text{Re}\, f_{ab}\right]^{-1} D^{a}D^{b},\label{VD1}
\ee
respectively, where $D^{a}=-g \frac{\partial G}{\partial \phi_{k}}(\tau^{a})_{k}^{l}\phi_{l}$ and $g$ is the gauge coupling constant corresponding to each gauge group and $\tau^{a}$ are corresponding generators. For $SU(2)_L$ symmetry $\tau^{a}=\sigma^{a}/2$, where $\sigma^{a}$ are Pauli matrices and the $U(1)_{Y}$ hypercharges of the fields $H_u$, $H_d$, $L$  are $Y = (\frac{1}{2}, -\frac{1}{2}, -\frac{1}{2})$ respectively. The quantity $f_{ab}$ is related to the kinetic energy of the gauge fields and is a holomorphic function of superfields $\phi_i$. We will consider a canonical form of $f_{ab}=\delta_{ab}$. The kinetic term of the scalar superfields is given by
\begin{equation}
\mathcal{L}_{KE}=K_{i}^{j*} \partial_{\mu}\phi^{i} \partial^{\mu}\phi^{*}_{j}~, 
\label{LK}
\ee
where $K^{i}_{j*}$ is the inverse of the K\"ahler metric $K_{i}^{j*} \equiv \partial^{2}K / \partial\phi^{i}\partial\phi^{*}_{j}$. 

 We will be assuming that the real part of $T$ field gets a vacuum expectation value
$<Re[T]> = c$ while the imaginary part $Im[T]=0$. The $vev$ of $T$ can be determined by 
some unspecified non-perturbative high-scale dynamics such as KKLT, KL or KL in Polonye model \cite{Dudas:2012wi,Linde:2011ja}.
Also, we assume that during inflation the charged fields take zero $vev$. For the above
considerations the $D$-term scalar potential is obtained as

\begin{equation}
V_{D}=\frac{9(g_{1}^{2}+g_{2}^{2})({\phi_{u}^{0}\phi_{u}^{0}}^{*} -{\phi_{d}^{0}\phi_{d}^{0}}^{*} + \widetilde{\nu_{L}} \widetilde{\nu_{L}^{*}})}{8(-3c + {\phi_{u}^{0}\phi_{u}^{0}}^{*} +{\phi_{d}^{0}\phi_{d}^{0}}^{*}-\widetilde{\nu_{L}} \widetilde{\nu_{L}^{*}} + \widetilde{N_{i}^{*}}\widetilde{N_{i}})^{2}}.
\end{equation}
For simplicity, we consider the following parametrization for the available
superfields as 

$\phi_{u}^{0} \to \phi \sin{\beta}$ , ${\phi_{u}^{0}}^{*} \to \phi_{s} \sin{\beta}$, $\phi_{d}^{0} \to \phi \cos{\beta}$, ${\phi_{d}^{0}}^* \to \phi_{s} \cos{\beta}$, $\widetilde{\nu_{L}} \to \gamma \phi$, ${\widetilde{\nu_{L}}^{*}} \to \gamma \phi_{s}$,  $\widetilde{N_{i} }\to n_{i} \phi$, ${\widetilde{N_{i}}^{*}} \to n_{i} \phi_{s}$,

For the above parametrization the kinetic term is obtained as 
\begin{equation}
\mathcal{L}_{KE} = \frac{9\; c\; (1+n_{2}^{2}+n_{3}^{2} + \gamma^{2})}{(-3\; c +(1+n_{2}^{2}+n_{3}^{2} + \gamma^{2}) |\phi|^{2})^{2}} |\partial_{\mu}\phi|^{2}
\end{equation}
we redefine the complex field $\phi_i$ to $\chi$ via
\bea
\phi=\frac{\sqrt{3c} \; \tanh[{\frac{\chi}{\sqrt{3}}}]}{\sqrt{(1+\gamma^2+n_{2}^{2}+n_{3}^{2})}}
\eea
With this field redefinition we get a canonical kinetic term for the vanishing imaginary
part of $\chi$ and its real part serves as the inflaton of the model
\begin{equation}
\mathcal{L}_{KE}= d\chi \; d\chi_{\ast} \operatorname{sech}\left[{\frac{\chi -\chi_{\ast}}{\sqrt{3}}}\right]^{2}
= |\partial_{\mu}\chi|^{2}
\end{equation}
The $D$-term potential in terms of inflaton $\chi$ becomes 
\begin{align}
    V_{D}=\frac{9(g_{1}^{2}+g_{2}^{2})(\gamma^{2} +\cos{[2 \beta]})^{2}}{8 (1+\gamma^{2}+n_{2}^{2}+ n_{3}^{2})^{2}} \sinh\left[\frac{\chi}{\sqrt{6}}\right]^{4}
\end{align}
It can be easily inferred from the above potential that a $D-$ flat direction for
inflation can be obtained for the condition $\gamma =\pm i \sqrt{\cos[2\beta]}$.

The $F$-term potential along the the observed $D$-flat direction is obtained as
\begin{align}
    V_{F} &= \frac{n_{3}^{2}\left(n^{2} M_{2}^{2} +M_{3}^{2}\right)}{4 c \left(n_{3}^{2}(1+n^{2})+2 \sin [\beta]^{2}\right)} \sinh\left(\sqrt{\frac{2}{3}} \chi\right)^{2} \left[1+\frac{C1}{A1}\tanh\left(\frac{\chi}{\sqrt{6}}\right)+ \frac{B1}{A1}\tanh\left(\frac{\chi}{\sqrt{6}}\right)^{2}\right] ,
    \label{vf}
\end{align}

where, we made a choice $ n_{2}=n\; n_{3}$ and the ratios $\frac{C1}{A1}$ and $\frac{B1}{A1}$
have the following form in terms of Yukawa couplings  $Y_{22}$ and $Y_{33}$, and heavy right handed neutrino masses $M_2$ and $M_3$ as 
\begin{align}
    \frac{B_1}{A_1}=\frac{2 i\sqrt{3}\;c\;\sqrt{\cos[2\beta]}\; \sin{[\beta](n M_{2}+y M_{3})Y_{22}}}{n_{3}(n^{2}M_{2}^{2}+M_{3}^{2})\sqrt{c(1+n_{3}^{2}(1+n^2)-\cos[2 \beta])}}
    \label{BA}
\end{align}
\begin{align}
    &\frac{C_1}{A_1}=\frac{3 c (1+y^2 +2 n_{3}^2 (n+y)^2 -2(1+y^2 +3 n_{3}^2 (n+y)^2)\cos[2 \beta]
    +(1+y^2)\cos[4 \beta]) Y_{22}^{2}}{4 n_{3}^{2}(1+n_{3}^{2}(1+n^2)-\cos[2 \beta]) (n^{2}M_{2}^{2}+M_{3}^{2})}\label{CA}\,,
\end{align}
where $y = Y_{33}/Y_{22}$. If we fix $\frac{C1}{A1}=1$ and $\frac{B1}{A1}=-2$, then the potential $V_F$ in  Eq. (\ref{vf})
becomes Starobinsky like given by
\begin{align}
    V_{F}(\chi)=\frac{1}{4}\lambda^{2} \;{\rm e}^{-\sqrt{\frac{2}{3}}\chi} \sinh \left[\frac{\chi}{\sqrt{6}}\right]^{2}
    \label{vfinf} 
\end{align}
where, 
\begin{align}
\lambda^{2}=\frac{3 n_{3}^{2}\left(n^{2} M_{2}^{2} +M_{3}^{2}\right)}{ c \left(n_{3}^{2}(1+n^{2})+2 \sin [\beta]^{2}\right)}
\label{lambdaCMB}
\end{align}
serves as the CMB normalization parameter. With the canonical kinetic term and scalar potential obtained in canonical inflaton field $\chi$,
the theory is now in the Einstein frame. Therefore, we can use the standard Einstein frame relations to estimate the 
inflationary observables, namely, amplitude of the curvature perturbation $\Delta_{\mathcal R}^{2}$, 
scalar spectral index $n_{s}$ and its running $\alpha_s$, and tensor-to-scalar ratio $r$, given by
\bea
\Delta_{\mathcal R}^{2} &=& \frac{1}{24 \pi^{2}} \frac{V_F}{\epsilon}\,,\label{amplitude}\\
n_{s} &=& 1-6\epsilon+2\eta\,,\\
\alpha_{s} &\equiv& \frac{dn_{s}}{d\ln k} = 16\epsilon\eta -24\epsilon^{2} - 2\xi\,,\\
r &=& 16\epsilon\,,
\eea
respectively. Here $\epsilon$, $\eta$ and $\xi$ are the 
slow-roll parameters, given by
\be
\epsilon = \frac{1}{2}\left(\frac{V_{F}'}{V_{F}}\right)^2, 
~~~~~~~\eta = \frac{V_{F}''}{V_{F}}\,,
~~~~~~~\xi = \frac{V_{F}'V_{F}'''}{V_{F}^{2}}.
\ee

For successful cosmology, it is required to have minimum $50-60$ $e$-folds of expansion during inflation. If we define $\chi_s$ and $\chi_e$ as the field values at the start and end of inflation respectively. Field $\chi_s$ value marks the point when observable CMB modes starts leaving the horizon and can be determined via the following relation
\be
N=\int_{\chi_f}^{\chi_i} \frac{V_F}{V_{F}'} d\chi\,,\label{efolds}
\ee 
and the end of inflation condition $\epsilon(\chi_e)=1$ fixes the field value $\chi_e$.

\begin{figure}
    \centering
    \begin{minipage}{0.46\textwidth}
        \centering
        \includegraphics[width=0.9\textwidth]{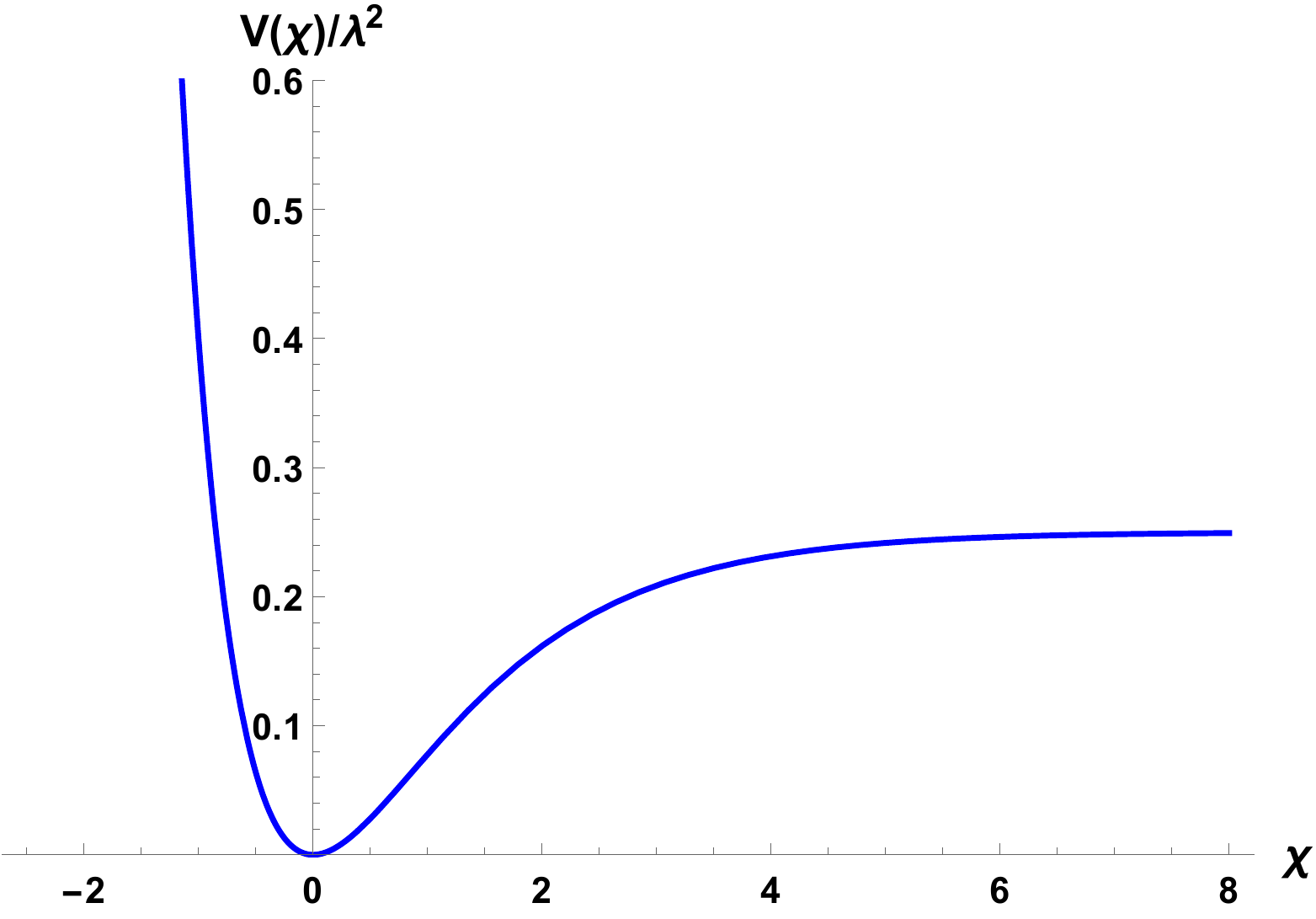} 
        \caption{\footnotesize $F$-term potential along the $D$-flat direction is shown.}
        \label{fig1}
    \end{minipage}\hfill
    \vspace{.4cm}
    \begin{minipage}{0.46\textwidth}
        \centering
        \includegraphics[width=0.9\textwidth]{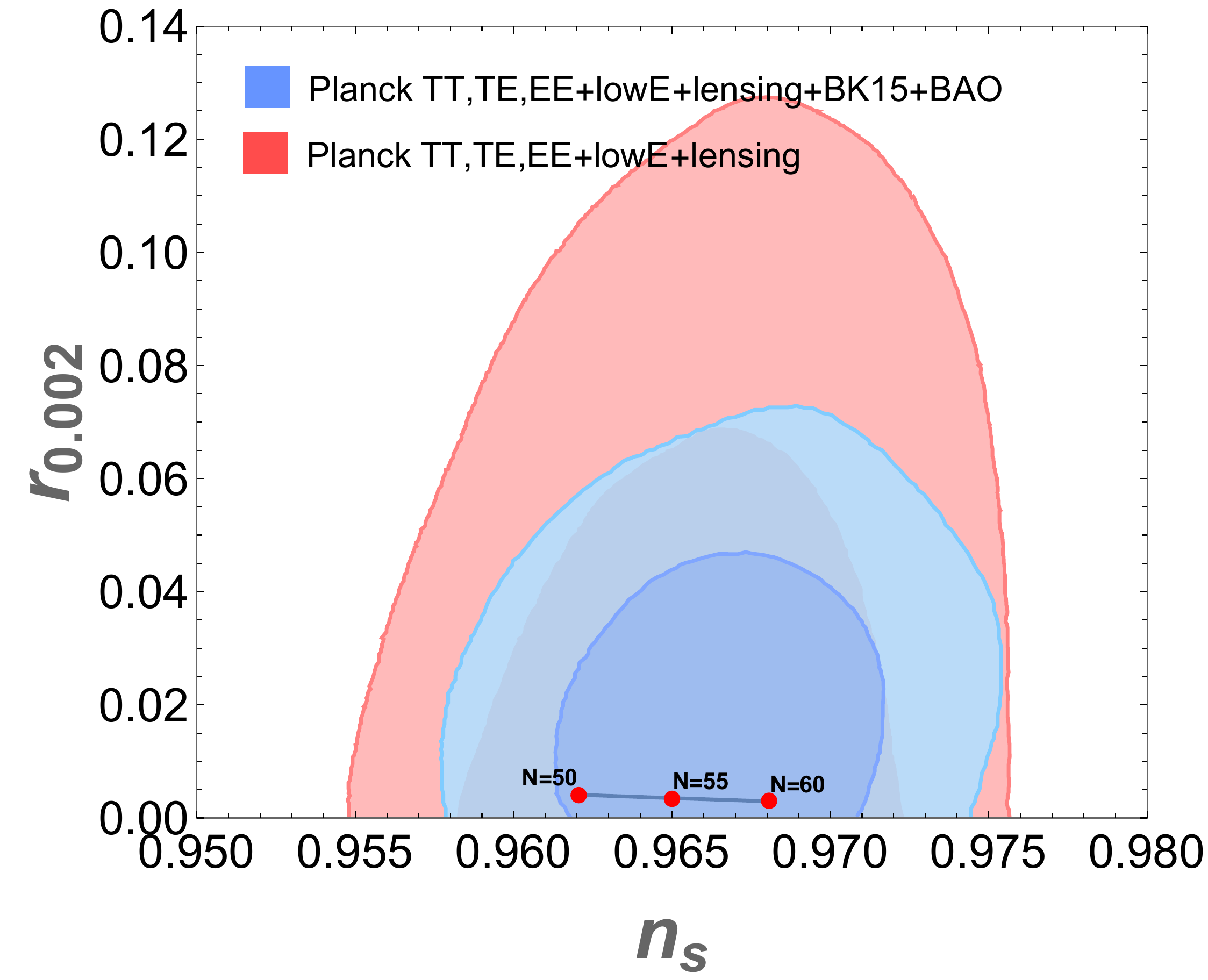} 
        \caption{\footnotesize In no-scale SUGRA Type-I Seesaw model, Starobinsky like $(r, n_s)$ model predictions are shown in the background of $1\sigma$ and $2\sigma$ regions as obtained by Planck-2018. }
        \label{fig2}
    \end{minipage}
    \vspace{.4cm}
\end{figure}

\subsection{Numerical analysis of the model}
\label{sec:NA}
From the Planck-2018 CMB temperature anisotropy data in combination with the $EE$ measurement at low multipoles and lensing (PlanckTT,TE,EE+lowE+lensing), we have the scalar amplitude, the spectral index and
its running as $\ln (10^{10} \Delta_{\mathcal R}^{2}) = 3.044\pm 0.014$, $n_{s}= 0.9649 \pm 0.0042$,
$\alpha_s = - 0.0045 \pm 0.0067$, respectively, at ($68 \%$ CL,)~ \cite{Akrami:2018odb}.
Also the Planck-2018 data combined with BK15 CMB polarization data
put an upper bound on tensor-to-scalar ratio $r_{0.002} < 0.056$ \,($95\%$ CL)~\cite{Ade:2018gkx}. 
Provided with observational results and having analytical results, we perform the numerical analysis of the model. We obtain $r_{ts} \simeq 0.003-0.004$, $n_s \simeq 0.963-0.967$ and $\alpha_s \simeq -(7.1-5.4)\times 10^{-4}$ for $50-60$ efold expansion as has been shown in the Fig.\ref{fig2}. The corresponding inflaton values are $\chi_s=5.44 - 5.28$ and CMB normalisation parameter $\lambda= (2.3 - 2)\times 10^{-5}$. The $F$-term potential along the $D$-flat direction is shown in Fig.\ref{fig1}. We may state now, that we have shown that a successful model of inflation in the no-scale supergravity in simple Type-I Seesaw model of Neutrino masses in NMSSM exists. We have derived our Starobinsky like model by imposing simple restrictions on Yukawa couplings $Y's$, Majorana mass $M's$ and other parameters of inflation potential in the form of $\frac{B1}{A1}$ eq.(\ref{BA}), $\frac{C1}{A1}$ eq.(\ref{CA}) and $\lambda^{2}$ eq.(\ref{lambdaCMB}) definitions discussed earlier. These conditions will be useful to unify the inflation model with the very important phenomenons of Reheating, Leptogenesis and light neutrino masses as we study in detail in next sections. 

\section{Leptogenesis in NMSSM} 
\label{sec:lepto}

For leptogenesis we will follow the earlier work on non-thermal soft leptogenesis\cite{Adhikari:2015ysa}. However as stated earlier we will consider this work in further details taking into account various scattering processes which could wash out asymmetry. From the detailed analysis using the Boltzmann equations, we will show the leptonic asymmetry and the freeze-out temperature for various choices of soft susy-breaking parameters and other parameters of the model.  For the type-I seesaw mechanism for generating light neutrino mass, the last two terms in Eq. \eqref{eq:Spot} are relevant and is written as 
\begin{eqnarray}
W_N & = & \frac{1}{2}M_i\hat{N_i^{c}}\hat{N_i^{c}}
+Y_{i\alpha}\hat{N_i^{c}}\hat{\ell}_{\alpha}\hat{H}_{u},
\end{eqnarray}
where $\hat{N_i^c}$, $\hat{\ell}_{\alpha}$ and $\hat{H}_{u}$ are the different chiral superfields as mentioned before.
Two heavier right-handed neutrino fields play important role in inflation. Leptogenesis occurs after the end of inflation. Particularly in supersymmetric theory, because of the production of gravitino and its' subsequent decays, to avoid the constraints from primordial nucleosynthesis, the reheating scale and the leptogenesis energy scale are expected to be below $10^9$ GeV. Because of that, we consider the masses $M_1$ and $M_2$ of the right handed neutrinos to be below $10^9$ GeV. $N_2$ could play role in reheating while  the lightest of the three heavy right-handed neutrinos $N_1$ with mass $M_1$ and its supersymmetric partner sneutrino are considered to play role in leptogenesis. In this section we will consider the notation
$N \equiv N_1$ and $Y_{\alpha} \equiv Y_{1\alpha}$.
The corresponding soft terms are
\begin{eqnarray}
-{\cal L}_{{\rm soft}} & = & \widetilde{M}^{2}\widetilde{N}^{*}\tilde{N}
+\left(\frac{1}{2}BM\widetilde{N}\widetilde{N}
+A_\alpha \widetilde{N}\widetilde{\ell}_{\alpha}H_{u}+{\rm H.c.}\right).
\end{eqnarray}

The mass and interaction terms involving the
sneutrino $\widetilde{N}$ from $W_{N}$ are given by
\begin{eqnarray}
-{\cal L}_{\widetilde{N}} & = & \left|M\right|^{2}\widetilde{N}^{*}\tilde{N}
+\left(M^{*}Y_{\alpha}\widetilde{N}^{*}\widetilde{\ell}_{\alpha}H_{u}
+Y_{\alpha}\overline{\widetilde{H}_{u}^{c}}P_{L}\ell_{\alpha}\widetilde{N}+{\rm H.c.}\right),
\end{eqnarray}
where $P_{L,R}=\frac{1}{2}\left(1\mp\gamma_{5}\right)$. 
Without loss of generality, the phases can be assigned only to the soft supersymmetry breaking parameter $A_\alpha$ without any loss of generality. We have considered generic
$A_\alpha$ and have not considered any relationship
$A_\alpha = A Y_\alpha$  as considered in
in Refs. \cite{D'Ambrosio:2003wy,Grossman:2003jv,Fong:2010qh}. The $A_\alpha$ 
couplings are found to give non-zero $CP$ violation  even at zero temperature.

Because of the bilinear $B$ term, $\widetilde{N}$ and $\widetilde{N}^{*}$ 
mix to form mass eigenstates
\begin{eqnarray}
\widetilde{N}_{+} & = & \frac{1}{\sqrt{2}}\left(\widetilde{N}+\widetilde{N}^{*}\right),\nonumber \\
\widetilde{N}_{-} & = & -\frac{i}{\sqrt{2}}\left(\widetilde{N}-\widetilde{N}^{*}\right),
\label{eq:mass_eigenstates}
\end{eqnarray}
with the corresponding masses given by 
\be
M_{\pm}^{2} =  M^{2}+\widetilde{M}^{2}\pm BM.
\label{eq:masses}
\ee
 The condition $B < M + \widetilde M^2/M$ is to be satisfied to avoid tachyonic masses with $\widetilde M < M$.  In the mass basis $\widetilde{N}_{\pm}$  the earlier Lagrangian can be written as:
\begin{eqnarray}
-{\cal L}_{\widetilde{N}}-{\cal L}_{{\rm soft}} & = & 
M_{+}^{2}\widetilde{N}_{+}^{*}\widetilde{N}_{+}+M_{-}^{2}\widetilde{N}_{-}^{*}\widetilde{N}_{-}\nonumber \\
 &  & +\frac{1}{\sqrt{2}}\left\{ \widetilde{N}_{+}\left[ Y_{\alpha} \overline{\widetilde{H}_{u}^{c}}P_{L}\ell_{\alpha}
+\left(A_\alpha + M Y_\alpha\right)\widetilde{\ell}_{\alpha}H_{u}\right]\right.\nonumber \\
 &  & \left.+i \widetilde{N}_{-}\left[ Y_{\alpha} \overline{\widetilde{H}_{u}^{c}}P_{L}\ell_{\alpha}
+\left(A_\alpha - M Y_\alpha\right)\widetilde{\ell}_{\alpha}H_{u}\right]+{\rm H.c.}\right\} .
\label{eq:lag}
\end{eqnarray}


For generation of leptonic asymmetry, Sakharov's basic three conditions are to be satisfied
\citep{sakh}:\\
 (1) There should be  baryon number ($B$) or lepton number ($L$) violating interactions and in our case it is possible through lepton number violating interactions in Eq. \eqref{eq:lag}\\
 (2) Both $C$ and $CP$ violating physical process as $CP$ phases comes from $A$ parameter in the decays of $\widetilde N_\pm$ and \\
 (3) Departure from the thermal equilibrium of those physical processes which means the decay width $\Gamma \lesssim H(T_f= M)$.
 where $H$ is the Hubble parameter. The total decay width $\Gamma_\pm $ for
 $\widetilde N_\pm$ is given by:
 \begin{align}
 \Gamma_\pm \approx \frac{M}{4\pi} \sum_\alpha \left[ Y_\alpha^2 + \frac{|A_\alpha|^2}{2 M^2} \pm \frac{Y_\alpha Re[A_\alpha]]}{M}\right]
 \label{eq:decaywidth}
 \end{align}
 in which to be in the perturbative regime, $A_\alpha, B < M$ and $Y_\alpha < 1$. The Hubble parameter $H$ is given by
 \be
 H= 1.66 \sqrt{g^*} T^2/M_{Pl} 
 \ee
 where $M_{Pl} = 1.22 \times 10^{19}$ GeV is the Planck mass and $g^* = 228.75$. So the out of equilibrium condition may be written as: 
 \begin{align}
     \sqrt{ \sum_\alpha\left[Y_\alpha^2 + \frac{|A_\alpha|^2}{2M^2} 
\pm \frac{Y_\alpha {\rm Re}(A_\alpha)}{M}\right] }
\lesssim 1.6 \times 10^{-5} \left(\frac{M}{10^7\,{\rm GeV}}\right)^{1/2}.
\label{eq:out-of-equilibrium}
 \end{align}
 
 All the above three conditions needs  to be fulfilled. But  for non-zero $CP$ asymmetry one requires interference among the  tree level and higher order Feynman diagrams related to those physical process. However, to get non-zero $B$ or $L$ asymmetry from the interference certain conditions  on  higher order Feynman diagram must be satisfied~ \cite{nano,raghu} which is that there should be $B$ or $L$ violation on the right of the cut on the on-shell internal lines of the higher order diagram. Later on we have considered only those diagrams in Figure \ref{feynplot} labelled as (1), (2) and (3) with continuous flow of lepton which satisfy the above criteria. Diagrams with reverse flow of lepton in the loop, will not contribute to leptonic asymmetry and have not been considered.
 \subsection{CP-Asymmetry}
 $\widetilde N_\pm$ decay to either to $\{\widetilde\ell_\alpha H_u\}$ or to $\{ \ell_\alpha \widetilde H_u\}$. Corresponding Feynman diagrams at tree level and higher order are shown in Fig. \ref{feynplot} based on the interactions in Eq.\eqref{eq:lag}. The higher order diagrams are the self energy correction, the interference between tree level and one-loop self energy diagrams is the source of $CP$ violation. Higher order vertex corrections diagram are not important as discussed in Ref.~\cite{Adhikari:2015ysa}.
\begin{figure}
    \centering
    \includegraphics[scale=0.55]{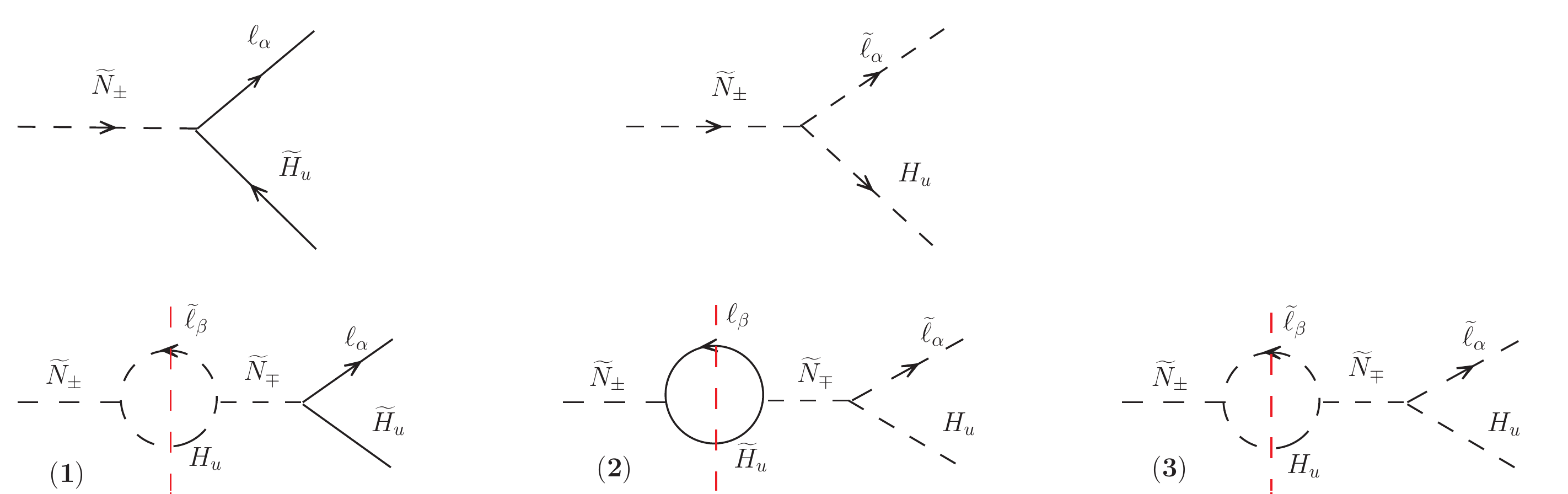}
    \caption{ Tree level diagram and one-loop self-energy correction diagrams for the decays 
$\widetilde N_\pm \to \ell_\alpha \widetilde H_u$ 
and $\widetilde N_\pm \to \widetilde\ell_\alpha H_u$. The direction of arrow in the loop indicates  flow of lepton number and the red  vertical lines shows the corresponding intermediate states go on mass shell.
The diagram  contribute to the $CP$ violation since they involve the soft couplings $A_\alpha$.}
    \label{feynplot}
\end{figure}
The $CP$ asymmetry parameter for the decays $\widetilde N_\pm \to \{\widetilde\ell_\alpha H_u, \ell_\alpha \widetilde H_u\}$ can be given as;
\begin{equation}
\epsilon^{S}_{\pm\alpha} \equiv 
\frac{\gamma(\widetilde N_\pm \to a_\alpha) 
- \gamma(\widetilde N_\pm \to \overline {a_\alpha})}
{\sum_{a_\beta;\beta}\left[\gamma(\widetilde N_\pm \to a_\beta) 
+ \gamma(\widetilde N_\pm \to \overline {a_\beta})\right]},
\label{eq:CP_asym}
\end{equation}
where $ a_\alpha \rightarrow \{\widetilde\ell_\alpha H_u, \ell_\alpha \widetilde H_u\}$ are different decay modes of sneutrino $\widetilde N $. The superscript $S$  indicates the asymmetry from  higher order self energy correction diagrams and $\overline {a_\alpha}$ is a $CP$ conjugate of $ a_{\alpha}$. $\gamma(\widetilde N_\pm \to a_\alpha$  is the thermally averaged  reaction rate for the process $\widetilde N_\pm \to a_\alpha$. The asymmetry calculation is done in the superequilibration regime, for which the temperature range is $T \lesssim 10^{8}$ GeV for $m_{susy} \sim \text{TeV}$~\cite{Fong:2010qh} and in this case, sleptons and leptons are indistinguishable (equal chemical potentials). Then for the leptonic asymmetry from particles and sparticles, the two  Boltzmann equations can be reduced  into the one single equation. Also it is possible to  sum over these CP asymmetries with leptons and sleptons in the final state.

 In the limit Ref.~\cite{Adhikari:2015ysa}
 \begin{align}
 Y_\alpha << | A_\alpha|/M,
 \label{YAM}
 \end{align}
 the non-thermal contribution is more than the thermal $CP$ asymmetry and also it is important to consider non-thermal one as it is not suppressed like thermal one because of $T$ dependence in thermal case through distribution function of the number densities of fermions and bosons. In the above limit 
\begin{eqnarray}
\epsilon^S_{\pm \alpha} 
& \simeq & \frac{1}{4\pi}\frac{|A_\alpha|^2}{\sum_\delta |A_\delta|^2}  
\sum_\beta Y_\beta \frac{{\rm Im} (A_\beta)}{M} 
\frac{4 B M}{4 B^2 + \Gamma_A^2},
\label{cpasy}
\end{eqnarray}
and the decay width Eq. \eqref{eq:decaywidth} can be approximated as :
\begin{align}
\Gamma_A \equiv \sum_\alpha \frac{|A_\alpha|^2}{8\pi M}.
\end{align}
The condition for resonance regime is
\begin{align}
\Gamma \approx B ,
\end{align}
for which  $\epsilon_{\pm} \sim (Y M)/|A|$.
Away from the resonance for $B >> \Gamma$ the $CP$ asymmetry $\epsilon_{\pm} \sim   10^{-1}Y|A|/B $ with the assumption that $CP$ phases are  $ {\cal O} (1)$. For  $|A| \sim$ TeV and B $>\Gamma$ with the out of  equilibrium condition, the $CP$ asymmetry is in the range $\epsilon_{\pm} \gtrsim 10^{-4-6}$ for $M \gtrsim 10^{6-7}$ GeV which may not be too low.
\subsection{Boltzmann Equations}
To calculate leptonic and hence baryonic asymmetry, apart from considering decays as discussed earlier, in the Boltzmann equations, we have taken into account various lepton number conserving and lepton number violating scattering processes which has effect on the number densities of $\widetilde N_+$ and $\widetilde N_-$ as shown in Eqs. \eqref{BEN+} and \eqref{BEN-}. Also we have considered lepton number violating scattering processes in Eq. \eqref{BENL} which could wash out the  lepton asymmetry generated through decays. 

To find out leptonic asymmetry, we have solved numerically all three coupled differential equations in  Eq. \eqref{BEN+}, \eqref{BEN-} and \eqref{BENL}, using the CP asymmetry parameter from the Eq. \eqref{cpasy}. 
\begin{eqnarray}
&& \frac{dY_{\widetilde N_+}(z) }{dz}=\frac{-1}{s(z)H(z)z}\bigg[\left(\frac{Y_{\widetilde N_+}}{Y^{eq}_{\widetilde N_+}}-1\right) \sum\limits_\alpha \bigg( \gamma_{ \widetilde N_+ \to \{\widetilde\ell_\alpha H_u \ell_\alpha \widetilde H_u\}}  + \sum\limits_\beta \left( \gamma_{\widetilde N_{+} \widetilde \ell^{\dagger}_\alpha \to \widetilde N^{\dagger}_{\pm} \widetilde \ell_\beta}+ \gamma_{\widetilde N_{+} \widetilde N^{\dagger}_{\pm}  \to  \widetilde \ell_\alpha \widetilde \ell^{\dagger}_\beta }\right)  \nonumber \\ 
&& +\quad \left. 3 \left( \gamma_{\widetilde N_{+}H_{u} \to \widetilde N^{\dagger}_{\pm} H_{u}^{\dagger}} + \gamma_{\widetilde N_{+} \widetilde N^{\dagger}_{\pm}  \to H_{u} H_{u}^{\dagger}}
 \right) \right) 
 + \left(\frac{Y_{\widetilde N_+}}{Y^{eq}_{\widetilde N_+}} \frac{Y_{\widetilde N_\pm}}{Y^{eq}_{\widetilde N_\pm}} -1\right) \sum\limits_\alpha \left( \sum\limits_\beta( \gamma_{\widetilde N_{+} \widetilde N_{\pm}  \to  \widetilde \ell_\alpha \widetilde \ell_\beta} )\right)
\bigg]
\label{BEN+}
\end{eqnarray}

\begin{eqnarray}
&&\frac{dY_{\widetilde N_-}(z) }{dz}=\frac{-1}{s(z)H(z)z}\bigg[\left(\frac{Y_{\widetilde N_-}}{Y^{eq}_{\widetilde N_-}}-1\right) \sum\limits_\alpha \bigg( \gamma_{ \widetilde N_- \to \{\widetilde\ell_\alpha H_u, \ell_\alpha \widetilde H_u\}}  + \sum\limits_\beta( \gamma_{\widetilde N_{-} \widetilde \ell^{\dagger}_\alpha \to \widetilde N^{\dagger}_{\pm} \widetilde \ell_\beta}+ \gamma_{\widetilde N_{-} \widetilde N^{\dagger}_{\pm}  \to  \widetilde \ell_\alpha \widetilde \ell^{\dagger}_\beta  }) \nonumber \\
&& +\quad \left. 3 \left( \gamma_{\widetilde N_{-}H_{u} \to \widetilde N^{\dagger}_{\pm} H_{u}^{\dagger}}+ \gamma_{\widetilde N_{-} \widetilde N^{\dagger}_{\pm}  \to H_{u} H_{u}^{\dagger}}\right) \right)
+ \left(\frac{Y_{\widetilde N_-}}{Y^{eq}_{\widetilde N_-}} \frac{Y_{\widetilde N_\pm}}{Y^{eq}_{\widetilde N_\pm}} -1\right) \sum\limits_\alpha \left( \sum\limits_\beta( \gamma_{\widetilde N_{-} \widetilde N_{\pm}  \to  \widetilde \ell_\alpha \widetilde \ell_\beta} )\right )\bigg]
\label{BEN-}
\end{eqnarray}

\begin{eqnarray}
\frac{dY_{\Delta L}(z) }{dz}&=&\frac{1}{s(z)H(z)z}\bigg[\left(\frac{Y_{\widetilde N_\pm}}{Y^{eq}_{\widetilde N_\pm}}-1\right) \sum\limits_\alpha \epsilon^S_{\pm \alpha}  \big(\gamma_{ \widetilde N_\pm \to \{\widetilde\ell_\alpha H_u, \ell_\alpha \widetilde H_u\}} \big) - \sum\limits_\alpha\frac{1}{2}\frac{Y_{\Delta L}}{Y_{ \widetilde\ell_{\alpha}}^{eq}} \big(\gamma_{ \widetilde N_\pm \to \{\widetilde\ell_\alpha H_u, \ell_\alpha \widetilde H_u\}} \big)\nonumber \\
&-& \sum\limits_{\alpha,\beta} \frac{Y_{\Delta L}}{Y_{ \widetilde\ell_{\alpha}}^{eq}} \bigg(\frac{Y_{\widetilde N_\pm}}{Y^{eq}_{\widetilde N_\pm}}\bigg) \big( \gamma_{\widetilde N_{\pm} \widetilde \ell^{\dagger}_\alpha \to \widetilde N^{\dagger}_{\pm} \widetilde \ell_\beta} \big) 
- \sum\limits_{\alpha,\beta} \frac{Y_{\Delta L}}{Y_{ \widetilde\ell_{\alpha}}^{eq}} \bigg(\frac{Y_{\widetilde N_\pm}}{Y^{eq}_{\widetilde N_\pm}} \frac{Y_{\widetilde N_\pm}}{Y^{eq}_{\widetilde N_\pm}} \bigg) \big( \gamma_{\widetilde N_{\pm}  \widetilde N_{\pm}  \to \widetilde \ell_\alpha  \widetilde \ell_\beta} \big)\bigg]
\label{BENL}
\end{eqnarray}
 where $Y_{\widetilde N_+} = n_{\widetilde N_+}/s $ , $Y_{\widetilde N_-} = n_{\widetilde N_-}/s $  and $Y_{\Delta L} = \frac{n_{\widetilde N}-n_{\bar{\widetilde N}}}{s} $ summed over for $\widetilde N_{\pm}$, $z = M/T$,  Hubble rate $H(z)=\sqrt{\frac{4\pi^3 g_*}{45}}\frac{M^{2}}{m_{pl}z^2}$ and  the entropy density $s(z) = g_{*}\frac{2\pi^2}{45}\frac{M^{3}}{z^3}$, with the effective number of degrees of freedom $g^{*} \sim 228.75$ in  MSSM.
For a decay the reaction rate $\gamma$ is given as \\
\begin{align*}
    \gamma_{\widetilde N_{\pm}} &= n^{eq}_{\widetilde N_{\pm}}\frac{K_1(z)}{K_2(z)}\bigg(\Gamma_{\widetilde N_{\pm}\rightarrow \widetilde \ell_{\alpha} H_{u}}\bigg)
\end{align*}
Here K$_1$ and K$_2$ are the modified Bessel functions and $\Gamma _{\widetilde {N}_{\pm}}$is the decay width of the decaying particle ${\widetilde{N}_{\pm}}$, and  superscript $eq$ denotes the corresponding number density of the particle when it is in thermal equilibrium. The reaction rate for a two body scattering in the above Boltzmann equations is evaluated from the below equation
\begin{align*}
    &\gamma_{a+b\rightarrow i +j} = \frac{M}{64\pi^4 z}\int^{\infty}_{s_{min}}ds \frac{2\lambda(s,m^2_{a},m^2_{b})}{s}\sigma(s)\sqrt{s}K_1\bigg(\frac{\sqrt{s}z}{M}\bigg) 
\end{align*}

Here $s$ is the squared centre of mass energy and 
\[ s_{min} = max[(m_{a}+m_{b})^2,(m_{i}+m_{j})^2]\]
\[\lambda (s,m_{a}^{2},m_{b}^{2}) \equiv [s-(m_{a}+m_{b})^2][s-(m_{a}-m_{b})^2]\]

The thermally averaged $\gamma$ which  corresponds to the scattering processes, their conjugate and inverse processes are not written separately because associated factors with $\gamma$ have been taken appropriately in Boltzmann equations.

Finally in presence of the sphalerons the generated leptonic asymmetry $Y_{\Delta L}$ gets converted to baryonic asymmetry  $Y_{\Delta B}$ as:
\begin{align*}
Y_{\Delta B} = \frac{n_{B}(z)-n_{\overline{B}}(z)}{s(z)} =  - \left( \frac{8 N_f + 4 N_H}{ 22 N_f + 13 N_H} \right) Y_{\Delta L}
\end{align*}
where $N_f =3$ is the total number of lepton  generations and $N_H =2$ is the total number of Higgs doublets in the  MSSM model. However,  this conversion  requires almost Ist order phase transition  \cite{saphos} which could be possible if the freeze out temperature $T_f$ is somewhat above electroweak scale and is  around 200 GeV or above. This is ensured from our plot on leptonic asymmetry $Y_{\Delta B}$ versus $z = M/T$ which has been shown later. However, before we go for leptonic asymmetry plot, we will discuss the constraints on $A$ and $B$ parameters which will be useful for that plot.

\subsection{Constraints on $A$ and $B$ parameters from neutrino mass, reheating and out of equilibrium condition}
\label{sec:results}
As the inflation mass scale is of the order of $M_3$ which is around $10^{13}$ GeV, whereas for reheating at the end of inflation we have considered it to be driven  by $M_2$ then to avoid gravitino problem, $M_2 \lesssim 10^9$ GeV. So $M_2 << M_3$. Now, under this condition, we are solving the two conditions of inflation in Eqs. \eqref{BA} and \eqref{CA} with $\frac{B_1}{A_1}=-2$ and $\frac{C_1}{A_1}=1$ then we obtain the solutions of two Yukawa couplings (which are present in the diagonal elements of the Dirac mass matrix block in the Type I seesaw mass matrix): 
\begin{align}
Y_{33}=\frac{2 i \lambda  M_{3}^2 \sin (\beta )}{\sqrt{\cos (2 \beta )} \left(3 M_{3}^2-c \lambda ^2 \left(n^2+1\right)\right)}    
\label{y13}
\end{align}
\begin{align}
Y_{22}=\frac{2 i \lambda ^2 M_{3}^2 \sin (\beta ) \left(c \lambda  n (3 \cos (2 \beta )-1)-\sqrt{\cos (2 \beta )} \sqrt{c (3 \cos (2 \beta )-1) \left(c \lambda ^2 \left(n^2+1\right)-3 M_{3}^2\right)}\right)}{\sqrt{\cos (2 \beta )} \left(c \lambda ^2 \left(n^2+1\right)-3 M_{3}^2\right) \left(c \lambda ^2 \left(\left(2 n^2-1\right) \cos (2 \beta )-n^2\right)+3 M_{3}^2 \cos (2 \beta )\right)}
    \label{y12}
\end{align}

These two expressions are essentially controlled by parameter $c \geq 1$ in $M_P =1$ units which has mass$^2$ dimension and very small parameter $\lambda \sim 10^{-5}$ from the normalization as discussed earlier. Other parameters like $n$, $n_3$ which are dimensionless does not vary much from 1. Because of 
these, even with some variations of other parameters, it is found that 
$Y_{22} \sim Y_{33}$. Now, three different light neutrino masses from Type I seesaw formula, can be written as $m_1 \approx Y_{11}^2 v^2/M_1$,  $m_2 \approx Y_{22}^2 v^2/M_2$ and  $m_3 \approx Y_{33}^2 v^2/M_3$ where $v$ is the $vev$ of the Higgs field. $M_1$ is supposed to be lesser than $M_2$ as it is controlling leptogenesis. Out of three light neutrino masses,  we cannot consider $m_2$ and $m_3$ to be heavier. This is because from neutrino oscillation experiments, they differ at most by an order but as $Y_{22} \sim Y_{33}$ and $M_2$ varies by several orders, it is not possible. Then we may consider $m_1 > m_2 > m_3$ and $m_3$ is lightest which may be considered to be almost zero or several order less than the other two, in the hierarchical light neutrino mass scenario. Under this scenario, setting $m_3 \approx 0$, one can estimate $m_1$ and $m_2$ from the two mass-squared splitting from neutrino oscillation experiments ~\cite{deSalas:2017kay,Capozzi:2018ubv,Esteban:2018azc} as $m_1 \approx 0.05$ eV and $m_2 \approx 0.009$ eV. Using the seesaw formula we can write
\begin{align}
    \frac{Y_{22}^2}{M_2} \approx 1.44 \times 10^{-16} \mbox{GeV}^{-1}
    \label{Y2M2}
\end{align}
and 
\begin{align}
    \frac{Y_{11}^2}{M_1} \approx 8.26 \times 10^{-16} \mbox{GeV}^{-1}
    \label{Y1M1}
\end{align}
where $M_1$ and $M_2$ are expressed in GeV units. So  like the $Y_{22} \sim Y_{33}$ as required by inflation, if we consider Yukawa couplings $Y_{11}$ to be of same order as $Y_{22}$, one can see that $M_2$ could be about one order higher than $M_1$. As leptogenesis is controlled by $M_1$ , $M_1 < M_2$ is required, as otherwise $\widetilde N_{2\pm} $ will control leptogenesis because the asymmetry generated by $M_1$ will be washed out. So we write $M_2 = k M_1$ where $k > 1$. Taking the ratio of the above two equations, we can write:

\begin{align}
    Y_{22} = 0.4175 \; {\sqrt{k} \; Y_{11}}
\end{align}
Using the non-thermal condition Eq.\eqref{YAM} for leptogenesis in $Y_{11}$ and using the relation $M_1 = M_2/k$ , we can write above equation as an inequality: 
\begin{align}
Y_{22} << 0.4175  \; k^{3/2} \; \frac{A_\alpha}{M_2}
\end{align}
 So for $k \approx 10$ with one order difference of $M_2$ with $M_1$, one may consider like the inequality for $M_1$, the inequality for $M_2$ as $Y_{22} < \frac{A_\alpha}{M_2} $. In general we can write the mass of $\widetilde N_{2\pm} $  approximately given by Eq. \eqref{eq:masses}, the decay width $\Gamma_{\widetilde N_{2\pm}}$  by Eq. \eqref{eq:decaywidth} after replacing $M$ by $M_2$ and replacing Yukawa coupling $Y_\alpha $ by $Y_{22}$ in Eq.~\eqref{eq:decaywidth}. Using the above inequality, we may simplify the decay width as :
\begin{align}
 \Gamma_{\widetilde N_{2\pm}} \approx \sum_\alpha \frac{|A_\alpha|^2}{8\pi \; M_2}  
 \label{eq:decayN2}
\end{align}

 Next we impose the non-thermal condition in \eqref{YAM} on \eqref{Y1M1} and obtain the inequality for $A_\alpha$ as:
\begin{equation}
    \frac{|A_\alpha|^2}{M_1^3} >>  8.26 \times 10^{-16} \mbox{GeV}^{-1}
    \label{AM1}
\end{equation}
where $|A_\alpha|$ is expressed in GeV units. 

Next, we discuss the condition related to reheating. As the universe cools down immensely at the end point of inflation, to reconcile with late time temperature, there should be some reheating phenomena at the end of inflation. We consider that the reheating is occurring due to the decay of $\widetilde N_{2\pm}$ with mass approximately given by\eqref{eq:masses} and decay width $\Gamma_{\widetilde N_{2\pm}}$ given by Eq. \eqref{eq:decaywidth} after replacing $M$ by $M_2$ in those equations respectively. The temperature $T_{rh}$ of the reheated Universe after inflation is given by : 
\begin{align}
     T_{rh} \approx \left(\frac{90}{\pi^2 geff}\right)^{(1/4)} \sqrt{\Gamma_{\widetilde N_{2\pm}} M_{Pl}}
     \label{eq:TrehM2}
\end{align}
 After replacing $\Gamma_{\widetilde N_{2\pm}}$ by \eqref{eq:decayN2} and $T_{rh} < 10^9$ GeV
 we can write
\begin{align}
\left(\frac{ 90}{\pi^2 g_{eff}}\right)^{1/4} \left(\frac{|A_\alpha|^2 M_{Pl}}{8 \pi k M_1}\right)^{1/2} < 10^9 \mbox{GeV}
\label{eq:trehM1}
\end{align}

where $M_1$ and $M_{Pl}$ are written in GeV unit. Using non-thermal condition \eqref{YAM}, the out of equilibrium condition \eqref{eq:out-of-equilibrium}  $i.e ~~ \Gamma _{\widetilde N_{1 \pm}} < H(T)$ and $H(T) \propto T^{2}$ we have put $T=M_{1}/z_{f}$
will be simplified to 
\begin{align}
|A_\alpha|  \lesssim 2.26274 \times 10^{-5}  \frac{M_1}{z_f} \left(\frac {M_1}{10^7 \mbox{GeV}}\right)^{1/2}
\label{eq:Afinal}
\end{align}
if the freeze out temperature $T_f$ differs from $M_1$ as $T_f =  M_1/z_f $. For higher $z_f$ there is more wash out from various scattering processes. On the other hand, higher $k$ values indicates bigger $M_2$ in comparison to $M_1$.  

Using the conditions in Eqs. \eqref{AM1}, \eqref{eq:trehM1} and \eqref{eq:Afinal} and for different values of $z_f$ and $k$ the allowed region in $A_\alpha$ and $M_1$ plane is shown in Figure \ref{AM2}. Particularly, we have checked for lower values of $z_f$, the allowed region increases. 
\begin{figure}
    \centering
    \includegraphics[scale=0.4]{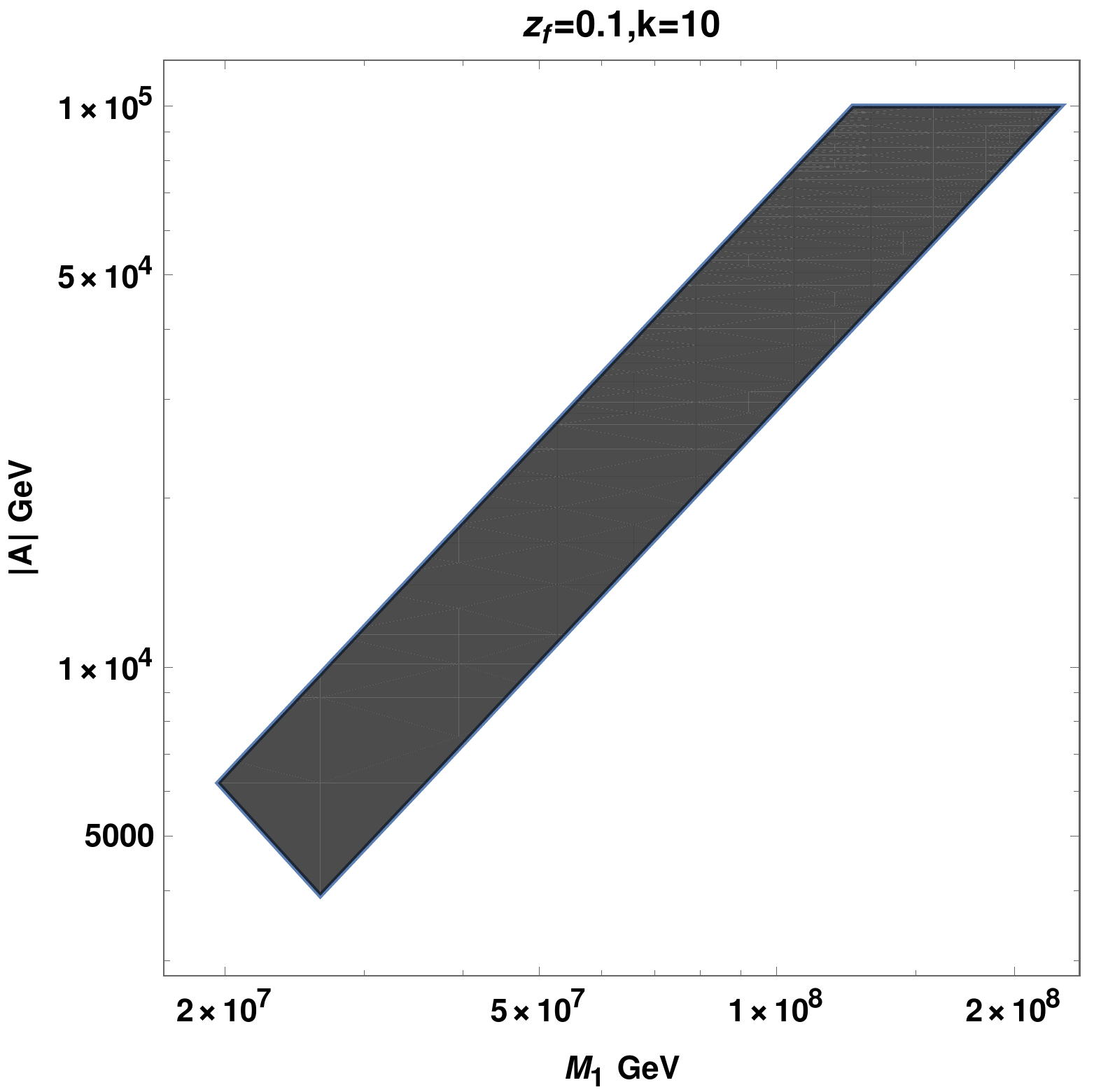}
     \includegraphics[scale=0.4]{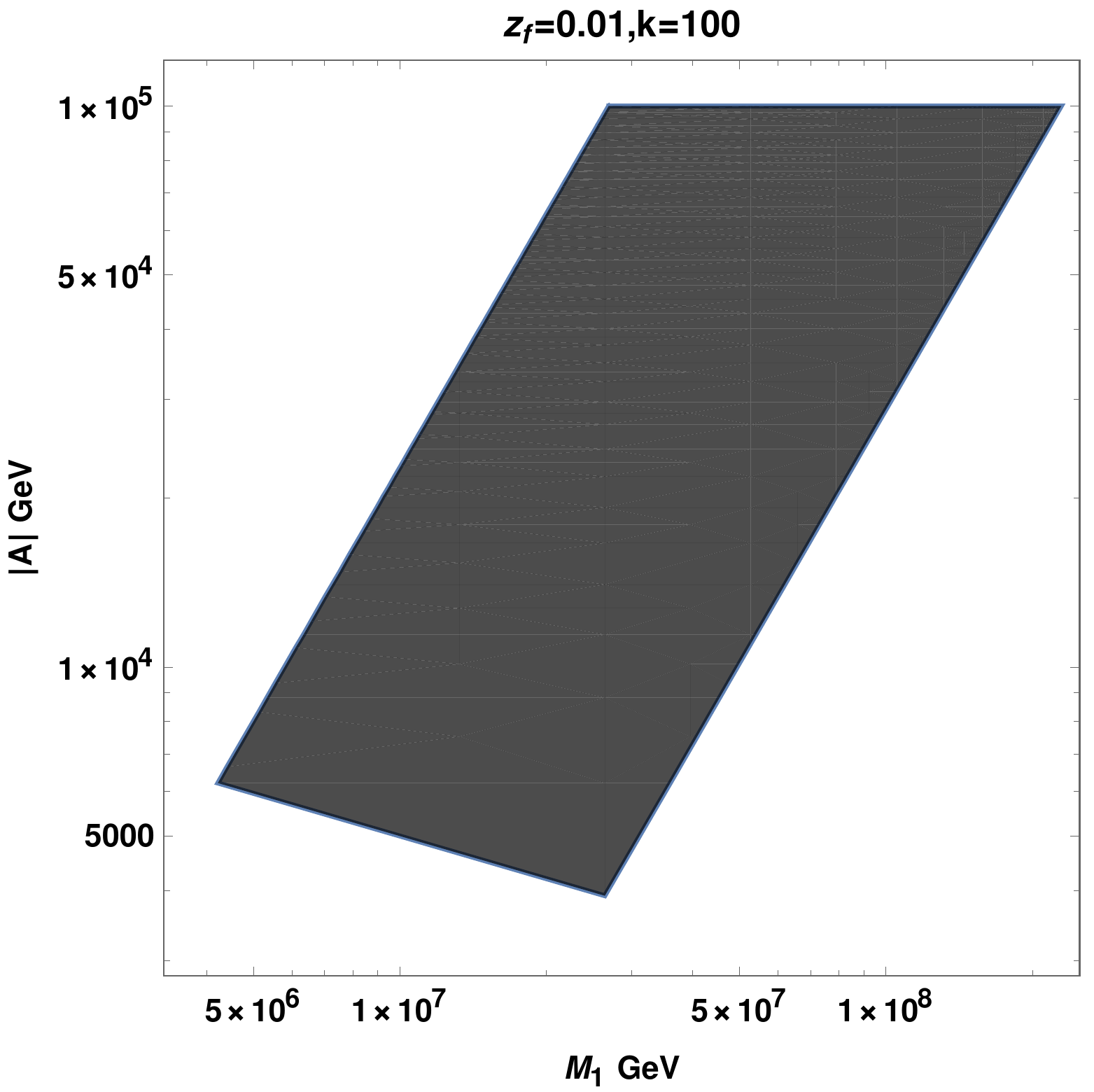}
    \caption{Allowed region of $A$ with the lightest right handed neutrino mass $M_1$. }
    \label{AM2}
\end{figure}
   
    In the resonance regime of leptogenesis, we have, $\Gamma_{\widetilde{ N}_{1\pm}} \approx B$.   Using this in eq. (4.27), the upper bound of $B$ is found as
\begin{align}
B \lesssim \frac{\pi g_{eff}^{1/2} k 10^{18}}{3\sqrt{10}M_{pl}} .
\end{align}
    in which $M_{pl}$ is in GeV unit.

   In plotting leptonic asymmetry in Figure \ref{asymmetryplot}, we have considered appropriate  values of $A_\alpha =A$  which are allowed for different values of $M_1$.
  After solving the Boltzmann Eqs. \eqref{BEN+}, \eqref{BEN-}
 and \eqref{BENL} numerically we have shown the evolution of $Y_{\Delta B}$ with temperature $(z= M_{ \widetilde {N}}/T)$ in Fig. \ref{asymmetryplot}. The observed baryonic asymmetry corresponds to  $Y_{\Delta B} \approx 10^{-11}$ based on recent experimental data ~\cite{Akrami:2018odb}  at recombination time. The required baryonic asymmetry has been obtained  around $z \approx 15 $, in which we have taken into account  the additional entropy dilution factor $f \approx 30$ \cite{early}. For the numerical analysis we have considered $M_{2}$  in the range of  $10^{6}$ to $10^8$ GeV. In the last  Boltzmann equation, there are  particularly two different kinds of scattering processes $\widetilde {N}_{\pm}  \widetilde {N}_{\pm}  \to \widetilde \ell_\alpha  \widetilde \ell_\beta$  and ${\widetilde N_{\pm} \widetilde \ell^{\dagger}_\alpha \to \widetilde N^{\dagger}_{\pm} \widetilde \ell_\beta}$  for which ${\Delta L} \neq 0$. As the lepton number violation of the decay process and that of the two scattering processes is same, so they enhance the asymmetry as shown in plot till $z \approx  0.3$ as dictated by  the kinematic and thermal factors of the Boltzmann equation. Later  the asymmetry starts dropping and finally reaches the freeze-out temperature  around $z \approx 15$. For the $A$, $B$ and $M_1$ parameters chosen by us, it is    found that  that the order of $B$ is nearer to $\Gamma$, that means we get the proper asymmetry near the resonance as discussed earlier. In this asymmetry plot both blue and orange curve corresponds nearer to resonance condition and suggest $B$ in GeV range. The black dotted line corresponds to the non-resonance condition. In this case, higher values of $B$ and $A$ could be possible.
   \begin{figure}{h}
    \centering
    \includegraphics[scale=0.5]{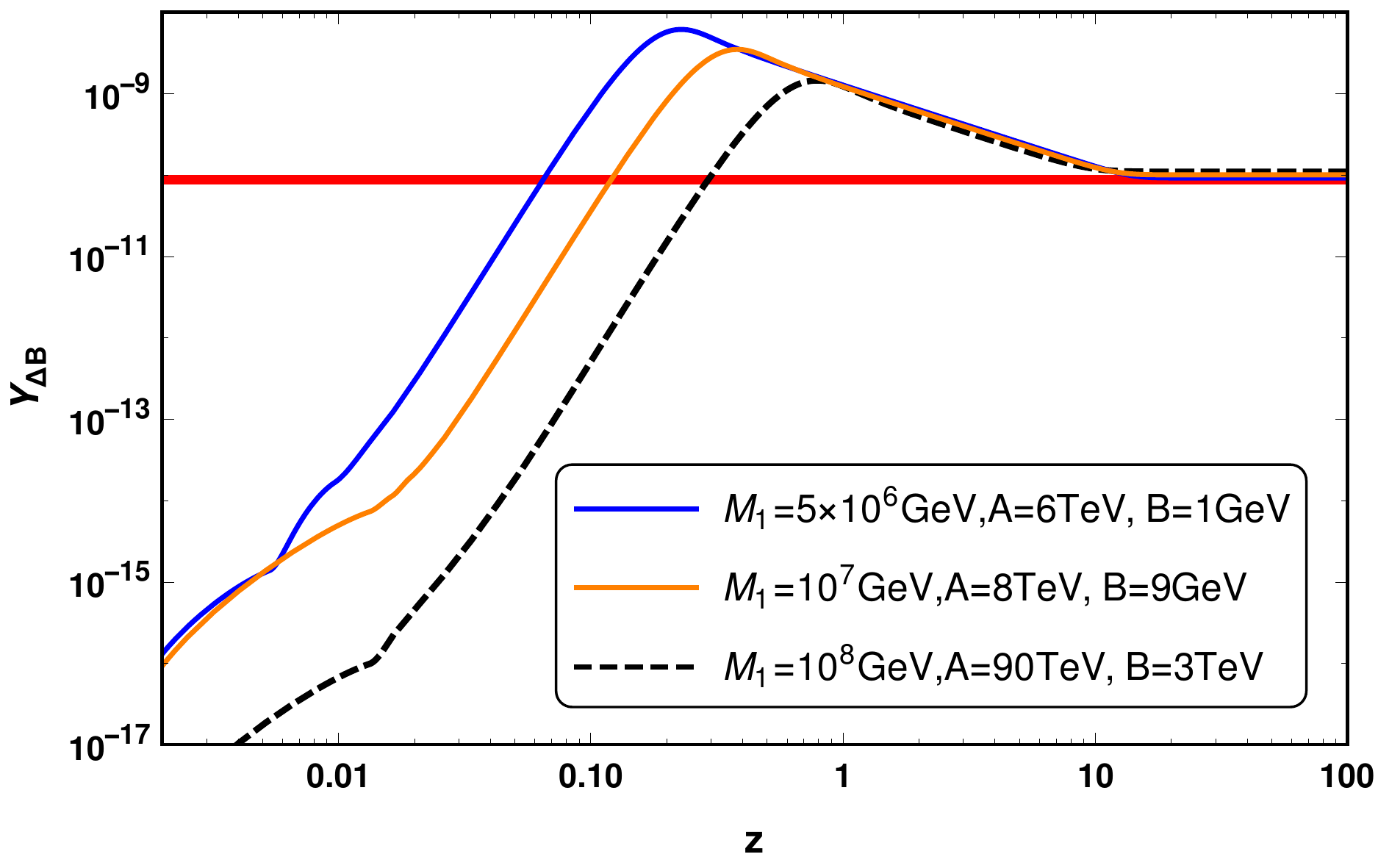}
    \caption{{ $Y_{\Delta B}$ versus $z=M/T$  for lightest Right Handed Neutrino mass  and soft breaking parameters $A ~~\text{and}~~B$ as mentioned in figure. }}
    \label{asymmetryplot}
\end{figure}
\section{Conclusions}
\label{sec:con}

In the supergravity embedded non-minimal supersymmetric standard model with three heavy right handed neutrino fields and their superpartners, i.e. in Type-I Seesaw superpotential, it is possible to have successful inflation, reheating and leptogenesis.  An appropriate choice of no-scale K\"ahler potential NMSSM superfields results in Starobinsky like inflation from the $F$-term potential where the inflation occurs along a sneutrino-Higgs $D$-flat direction and gives appropriately different cosmological parameters as observed.

In Type-I Seesaw framework of active neutrino masses,  the heavier two have been considered for inflation. In keeping the amplitude consistent with observation, the approximate mass scale for the heaviest right handed neutrino is obtained around $10^{13}$ GeV  as the other parameters in the expression of $\lambda$ have been assumed not to vary much from 1.  As the  decay of $M_2$ has been considered for reheating, the mass scale of it is expected to be below $10^9$ GeV which is several order below $M_3$. The leptonic asymmetry and the leptogenesis mechanism is controlled by $M_1$ and after numerically solving Boltzmann equations and taking into account various scattering processes, it is found that it could be in the approximate range of $10^6$ to $10^8$ GeV. In getting the Starobinsky like potential from $V_F$ potential with D-flat direction  for inflation, two Yukawa couplings $Y_{23}$ and $Y_{33}$ are required to be of almost same order. Then from seesaw relations of heavy right handed neutrino, it is found that the lightest active neutrino mass should be mainly related with $M_3$, while the other light  neutrino mass  which is slightly heavier than the earlier one, should mainly depend on $M_2$. While the heaviest one among three light neutrino masses should depend mainly on $M_1$. 
  
  The soft susy breaking  $A$ and $B$ parameters have played role in leptogenesis. Using the non-thermal condition in soft leptogenesis, the out of equilibrium condition and also using the upper bound on reheating temperature, the allowed region in $A$ and $M_1$ plane has been found. Using resonance condition in soft leptogenesis as well as the condition on reheating temperature, the constraint on $B$ parameter has been obtained. Constraining these soft parameters  could give us some understanding of the hidden sector spontaneous symmetry breaking in Supergravity as well as signify the low-energy phenomenology of Non-minimal Supersymmetric Standard model.


\section*{Acknowledgements}
  I. A. B would like to thank Department of Science and
Technology (DST), Govt. of India for financial support
through Inspire Fellowship [Department of Science and
Technology, Ministry of Science and Technology, (DST/
INSPIRE Fellowship/2014/IF140038), (Republic of India/
IN)]. I.A.B. would also like to thank Yogesh Jangra for useful discussions on cosmological observables. GKC would like to thank University Grants Commission, Govt. of India to provide financial support via Dr. D. S. Kothari Postdoctoral Fellowship (Grant No. BSR/PH/2017-18/0026). G.K.C. also would like to thank Prof. Subhendra Mohanty and Dr. Najimuddin Khan for useful discussions. 
\\
\providecommand{\href}[2]{#2}\begingroup\raggedright\endgroup

\end{document}